\documentclass[aps,preprint,nofootinbib,preprintnumbers,eqsecnum]{revtex4-1}

\usepackage{epstopdf}
\usepackage{color}

\newcommand{\nb}[1]{\color{blue}}

\newcommand{\hl}[1]{\color{magenta}}

\usepackage[
	  pagebackref=false,
	  colorlinks=true,
      linkcolor=blue,
      urlcolor=blue,
      filecolor=black,
      citecolor=red,
      pdfstartview=FitV,
      pdftitle={},
        pdfauthor={Michael Crossley, Paolo Glorioso, Hong Liu, Yifan Wang},
        pdfsubject={},
        pdfkeywords={},
        pdfpagemode=None,
        bookmarksopen=true
      ]{hyperref}

\usepackage[normalem]{ulem}
\usepackage{amsmath}
\usepackage{enumerate}
\usepackage{amsfonts}
\usepackage{epsfig}

\def\Tr{\mathop{\rm Tr}}
\def\tr{\mathop{\rm tr}}

\newcommand\half{{\ensuremath{\frac{1}{2}}}}
\newcommand\p{\ensuremath{\partial}}

\newcommand\field[1]{{\ensuremath{\mathbb{{#1}}}}}

\newcommand{\KK}{\field{K}}

\newcommand{\be}{\begin{equation}}
\newcommand{\ee}{\end{equation}}
\newcommand{\bea}{\begin{eqnarray}}
\newcommand{\eea}{\end{eqnarray}}

\newcommand{\bi}{\begin{itemize}}
\newcommand{\ei}{\end{itemize}}
\newcommand{\ben}{\begin{enumerate}}
\newcommand{\een}{\end{enumerate}}
\newcommand{\bca}{\begin{cases}}
\newcommand{\eca}{\end{cases}}
\newcommand{\bln}{\begin{align}}
\newcommand{\eln}{\end{align}}
\newcommand{\bst}{\begin{split}}
\newcommand{\est}{\end{split}}

\newcommand\al{{\alpha}}
\newcommand\ep{\epsilon}
\newcommand\sig{\sigma}
\newcommand\Sig{\Sigma}
\newcommand\lam{\lambda}
\newcommand\Lam{\Lambda}

\newcommand\ga{{\ensuremath{{\gamma}}}}

\newcommand\de{{\ensuremath{{\delta}}}}
\newcommand\De{{\ensuremath{{\Delta}}}}

\newcommand\nab{{\nabla}}

\newcommand\ov{\over}
\newcommand\ha{{\half}}

\def\le{\left}
\def\ri{\right}

\newcommand\sD{{\ensuremath{{\mathcal D}}}}

\newcommand\sK{{\ensuremath{{\mathcal K}}}}
\newcommand\sL{{\ensuremath{{\mathcal L}}}}
\newcommand\sM{{\ensuremath{{\mathcal M}}}}

\newcommand\sR{{\mathcal R}}

\newcommand{\sdet}{{\rm Sdet}}

\newcommand{\fn}{{\mathfrak n}}

\newcommand{\suv}{{\Sig_{\rm UV}}}
\newcommand{\sir}{{\Sig_{\rm IR}}}

\newcommand{\rmi}{{\rm i}}
\newcommand{\rmj}{{\rm j}}
\newcommand{\rmk}{{\rm k}}

\newcommand{\dR}{{^{(d)} \!R}}

\begin{document}

\title{Off-shell hydrodynamics from holography}

\preprint{MIT-CTP/4668}

\author{Michael Crossley, Paolo Glorioso, Hong Liu and Yifan Wang}
\affiliation{Center for Theoretical Physics, \\
Massachusetts
Institute of Technology,
Cambridge, MA 02139 }

\begin{abstract}

\noindent
We outline a program for obtaining an action principle for dissipative fluid dynamics by considering the holographic Wilsonian renormalization group applied to systems with a gravity dual.
As a first step, in this paper we restrict to systems with a non-dissipative horizon. By integrating out gapped degrees of freedom in the bulk gravitational system between an asymptotic boundary and a horizon, we are led to a
formulation of hydrodynamics where the dynamical variables are not standard velocity and
temperature fields, but the relative embedding of the boundary and horizon hypersurfaces.
 At zeroth order, this action reduces to that proposed by Dubovsky et al. as an off-shell formulation of ideal fluid dynamics.

\end{abstract}

\today

\maketitle

\tableofcontents

\section{Introduction}

At distance and time scales much larger than the inverse temperature and any other
microscopic dynamical scales, a quantum many-body system in local thermal equilibrium
should be described by hydrodynamics.  Except for ideal fluids, the current formulation of hydrodynamics has been on the
level of equations of motion. There are, however, many physical situations where hydrodynamical fluctuations
play an important role. An action principle is greatly desired. There are two main difficulties. One is to properly treat dissipation, and the other is to find the right set of dynamical degrees of freedom to formulate an action principle, as standard variables such as the velocity field appear not suitable.

In principle it should be possible to derive hydrodynamics as a low energy effective field theory from a quantum field theory  at a finite temperature via Wilsonian renormalization group (RG) by integrating out all gapped modes, but
in practice it has not been possible to do so. Such a formulation should lead to an action principle.

 For holographic systems,
the holographic duality \cite{Maldacena:1997re,Gubser:1998bc,Witten:1998qj} provides a striking geometric description of the
renormalization group flow in terms of the radial flow in the bulk geometry.
In particular, the Wilsonian renormalization group flow of a boundary
system can be described on the gravity side by integrating out part of the bulk
spacetime along the radial direction~\cite{Heemskerk:2010hk,Faulkner:2010jy}. The proposal expresses the Wilsonian effective action in terms of
a gravitational action defined at the boundary of the remaining spacetime region.

\begin{figure}[!h]
\begin{center}
$
\begin{array}{cc}
\includegraphics[scale=0.75]{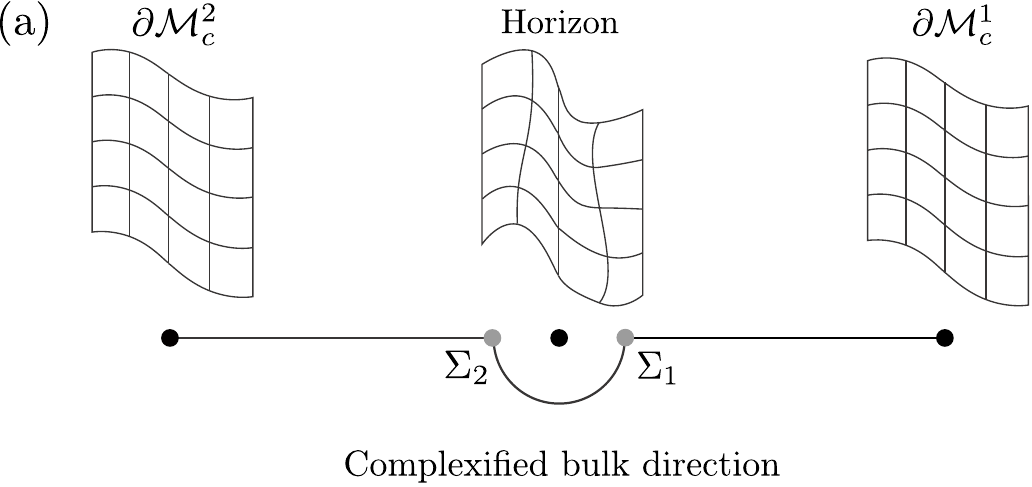} \,\,\,\,\,\,\,\,\,\,\,\, &
\includegraphics[scale=1]{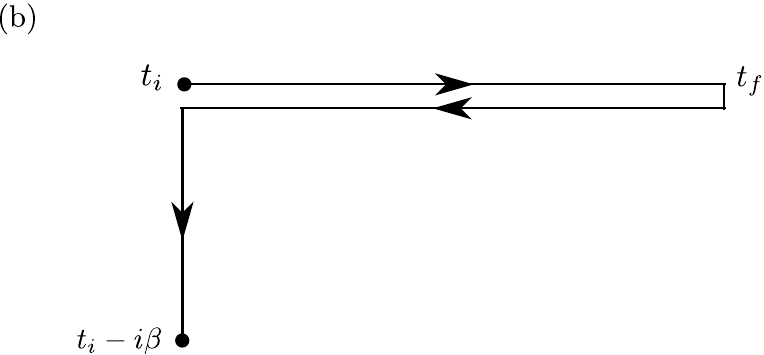}
\end{array}
$
\end{center}
\caption{(a) Complex bulk manifold $\sM_c$ consisting of two copies of asymptotic AdS spacetimes patched together at a  horizon hypersurface. Also labeled are stretched horizons $\Sig_1, \Sig_2$ discussed around~\eqref{1.2}.
(b) A boundary theory Schwinger-Keldysh contour used to describe non-equilibrium physics.
The two AdS regions map to the two horizontal legs of the Schwinger-Keldysh contour, while the analytic continuation around the horizon corresponds to the vertical leg which defines the initial thermal density matrix.
 }
 \label{fig:SK}
\end{figure}

The purpose of the current paper is to take a first step toward deriving an action for hydrodynamics using holographic Wilsonian RG.\footnote{The connections between 
holography and hydrodynamics has by now quite a history, starting with~\cite{Policastro:2001yc,Policastro:2002se,Policastro:2002tn,Kovtun:2003wp} and 
culminated in the fluid/gravity correspondence~\cite{Bhattacharyya:2008jc,Rangamani:2009xk,Hubeny:2011hd}.} 
The basic idea is as follows: consider the gravity path integral
\be \label{pane}
Z [\bar g_1, \bar g_2] = \int_{\sM_c} D G \, e^{i S[G]}
\ee
over metrics $G$ on a complex bulk manifold $\sM_c$ consisting of two copies of asymptotic AdS spacetimes patched together at a {\it dynamical} horizon hypersurface, as shown in Fig.~\ref{fig:SK}. $\sM_c$ has two asymptotic boundaries $\p \sM_c^{1,2}$ with boundary metrics $\bar g_1, \bar g_2$ respectively. The horizon is dynamical as its metric is integrated over.  The two copies of AdS can be considered as corresponding to the two long horizontal legs of a Schwinger-Keldysh contour, with the continuation around the horizon corresponding to the vertical leg~\cite{Herzog:2002pc}.  
In~\eqref{pane} one integrates out all gapped degrees of freedom, and the resulting effective action for whatever gapless degrees of freedom remain is the desired action for hydrodynamics. For this purpose, it is convenient to introduce stretched horizons $\Sig_\al, \al =1,2$ on each slice of the bulk manifold, which separate the bulk manifold
into three different regions (see Fig.~\ref{fig:SK}), i.e.
\be \label{1.2}
\int_{\sM_c}  = \int_{\p \sM_c^1}^{\Sig_1}  + \int_{\Sig_1}^{\Sig_2}  + \int_{\Sig_2}^{\p \sM_c^2}
\ .
\ee
The bulk path integral can be written as
\be \label{decpl}
Z [\bar g_1, \bar g_2]= \int_{\Sig_1} D\bar h_1 \int_{\Sig_2} D\bar h_2    \, \Psi_{\rm IR} [\bar h_1, \bar h_2] \,
\Psi_{\rm UV} [\bar h_1, \bar g_1] \Psi_{\rm UV}^* [\bar h_2, \bar g_2]
\ee
where $\bar h_{1}$ and $\bar h_2$ are induced metric on the stretched horizons. Various factors in the integrand of~\eqref{decpl} arise from the path integrals in three regions, e.g.
\be\label{uvint}
\Psi_{\rm UV} [\bar h_1, \bar g_1] = \int_{\bar g_1}^{\bar h_1} D G \, e^{i S[G]}
\ee
integrates over all metrics $G$ between $\p \sM_c^1$ and $\Sig_1$ with Dirichlet boundary conditions $\bar g_1$ and $\bar h_1$, and similarly with the others.
The complex conjugate on $\Psi_{\rm UV}^* [\bar h_2, \bar g_2]$ is due to that the bulk manifold in the region between $\Sig_2$ and $\p \sM_c^2$ has the opposite orientation from that between $\p \sM_c^1$ and $\Sig_1$.

Connections between hydrodynamics and Schwinger-Keldysh contour have been made recently in various contexts in~\cite{Son:2009vu,CaronHuot:2011dr,Grozdanov:2013dba,Haehl:2013hoa,Haehl:2014zda,Grozdanov:2015nea,Harder:2015nxa,Haehl:2015pja}. 

\begin{figure}[!h]
\begin{center}
\includegraphics[scale=1]{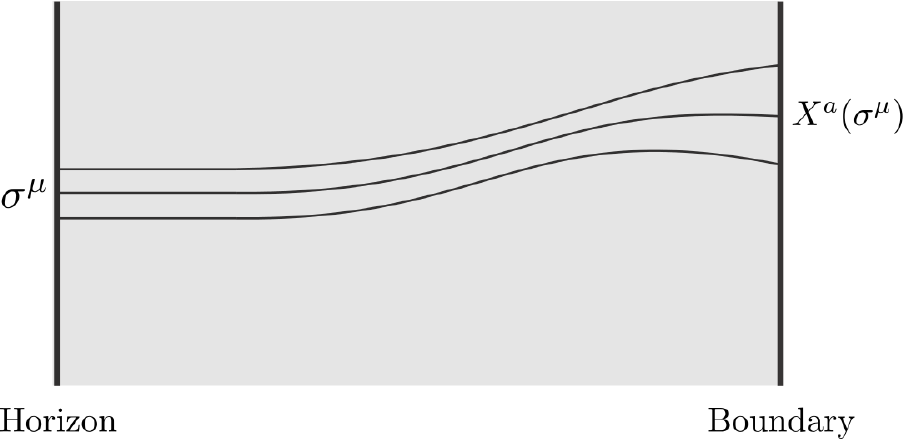}
\end{center}
\caption{The gapless degrees of freedom in the path integrals~\eqref{uvint} are the relative embedding coordinates $X^a_1 (\sig^\mu)$ between the horizon and the boundary. $X^a_1$ can be understood geometrically as follows: start with $\sig^\mu$ at $\Sig_1$, shooting a congruence of geodesics orthogonal to $\Sig_1$ toward the boundary, the intersections of these geodesics with the boundary define $X^a_1$.
 }
 \label{fig:embed}
\end{figure}

 In this paper we describe integrations over {\it gapped} degrees of freedom in the path integral~\eqref{uvint}. As anticipated earlier by Nickel and Son~\cite{Nickel:2010pr}, in~\eqref{uvint} the only gapless degrees of freedom are
the relative embedding coordinates $X^a_1 (\sig^\mu)$ of the boundary $\sM_c^1$ and the stretched horizon hypersurface $\Sig_1$, see
Fig.~\ref{fig:embed}. Integrating out all other degrees of freedom we obtain an effective action $I_{\rm UV} [X^a_1, \bar g_1, \bar h_1]$ for embeddings $X^a_1$, i.e.~\eqref{uvint} becomes
\be\label{uvint1}
\Psi_{\rm UV} [\bar h_1, \bar g_1] = \int DX^a_1 \, e^{i I_{\rm UV} [X^a_1, \bar g_1, \bar h_1]} \ .
 \ee
We develop techniques to compute $I_{\rm UV} \left[X^a_1, \bar g_1, \bar h_1\right]$ in expansion of boundary derivatives  at full nonlinear level in a saddle point approximation.
 Plugging~\eqref{uvint1} into~\eqref{decpl} and evaluating $\bar h_1, \bar h_2$ integrals one then obtains the full hydrodynamical action in terms of $X_1^a$ and $X_2^a$, i.e.
 \be
 Z [\bar g_1, \bar g_2] = \int DX_1^a DX_2^a \, e^{i I_{\rm hydro} [X_1, X_2; \bar g_1, \bar g_2]}
 \ee
with
\be \label{enne}
e^{i I_{\rm hydro} [X_1, X_2; \bar g_1, \bar g_2]} = \int_{\Sig_1} D\bar h_1 \int_{\Sig_2} D\bar h_2    \, \Psi_{\rm IR} [\bar h_1, \bar h_2] \, e^{i (I_{\rm UV} [X^a_1, \bar g_1, \bar h_1] - I_{\rm UV} [X^a_2, \bar g_2, \bar h_2])} \ .
\ee
  The evaluation of $\Psi_{\rm IR} \left[\bar h_1, \bar h_2\right]$ requires developing new techniques for analytic continuations through the horizon. We will leave its discussion and the full evaluation of~\eqref{enne} elsewhere.
Hydrodynamical actions based on doubled $X^a$ degrees of freedom discussed here have also been discussed 
recently in~\cite{Grozdanov:2013dba,Haehl:2013hoa,Haehl:2014zda,Grozdanov:2015nea,Harder:2015nxa,Haehl:2015pja}.

We also show that at zeroth order in the derivative expansion, if one (i) takes $\bar h_1$ to the horizon, i.e. making $\Sig_1$ a null hypersurface, and (ii) requires $\bar h_1$ to be non-dissipative, i.e. the local area element is constant along
the null geodesics  which generate the horizon, $\bar h_1$ completely decouples from
$I_{\rm UV} [X_1^a, \bar g_1, \bar h_1]$, and $I_{\rm UV}$ reduces to the conformal version of the ideal fluid action proposed by Dubovsky et al~\cite{Dubovsky:2005xd,Dubovsky:2011sj},~i.e.
\be
I_{\rm UV} [X_1^a, \bar g_1, \bar h_1] = I_{\rm ideal} [\xi_1, \bar g_1]
\ee
where
\be \label{dubo1}
 I_{\rm ideal} [\xi, \bar g]  = - (d-1) \int d^d \sig  \sqrt{-\bar g}  \, \le(\det \alpha^{-1} \ri)^{d \ov 2 (d-1)},~~(\alpha^{-1})^{\rmi \rmj}=\bar g^{\mu\nu}   \partial_\mu \xi^{\rmi}  \partial_\nu  \xi^{\rmj},
\ee
and $\xi^{\rmi} (\sig^\mu)$ (with $\rmi = 1, 2, \cdots, d-1$) are embeddings $X^a (\sig^\mu)$ for a null hypersurface for which the time direction decouples. In particular, the volume-preserving diffeomorphisms which played a key role in the formulation of~\cite{Dubovsky:2005xd} arise here as residual freedom of horizon diffeomorphism. The entropy current also arises naturally as the Hodge dual of the pull-back of the horizon area form to the boundary.

It is tempting to ask whether conditions (i) and (ii) in the previous paragraph will also lead to a non-dissipative fluid action at higher orders. We find, however, that the 2nd order action is divergent
unless one is restricted to shear-free flows. While it makes sense to make such restrictions in an equation of motion,
imposing it at the level of path integrals for $X^a$ appears problematic. We thus conclude that one must include dissipation
in order to have a consistent formulation.

We also note that the fact we find~\eqref{dubo1} when pushing $\Sig_1$ to the horizon does {\it not} necessarily imply that at
zeroth order  the full effective action~\eqref{enne} will be given by
\be
I_{\rm hydro} =  I_{\rm ideal} [\xi_1, \bar g_1] -  I_{\rm ideal} [\xi_2, \bar g_2]
\ee
as the integrations over $\bar h_1, \bar h_2$ will generate new structures.
At this stage the precise relation between~\eqref{dubo1} and the zeroth order of $I_{\rm hydro}$ is not yet fully clear to us.

The plan of the paper is as follows. We will explain our holographic setup and the gravitational boundary value problem in details in Sec.~\ref{sec:ii}. In Sec.~\ref{thirdsec}, we perform the path integral~\eqref{uvint} using saddle point approximation to obtain $I_{\rm UV}$ defined in~\eqref{uvint1} at zeroth order in the boundary derivative expansion and relate it
to the ideal fluid action~\eqref{dubo1}. In Sec.~\ref{sec:2nd} we briefly comment on the generalization to higher orders in the derivative expansion. We conclude with a discussion of open questions and future directions in Sec.~\ref{sec:diss}.

{\it Note added:} We understand Jan de Boer, Natalia Pinzani Fokeeva and Michal Heller have obtained similar
results~\cite{deBoer:2015ija}.

\section{Setup} \label{sec:ii}

In this section, we describe in detail our setup for computing~\eqref{uvint} to obtain $I_{\rm UV} [X^a, \bar g, \bar h]$
defined in~\eqref{uvint1}.

\subsection{Isolating hydrodynamical degrees of freedom}

In this subsection, we describe a series of formal manipulations of path integrals for gravity which allow
us to isolate $X^a$ as the ``would-be'' hydrodynamical degrees of freedom. There are standard difficulties in defining rigorously path integrals for gravity, which will not concern us as we will be only interested in the path integrals at a semi-classical level, i.e. in terms of saddle points and fluctuations around them.

Consider a path integral of the form
\be \label{path}
\Psi [\bar h, \bar g]  
= \int_{\bar g}^{\bar h} D G \, e^{{i } S[G]}
\ee
where the integration is over all spacetime metrics
\be \label{met0}
ds^2 = G_{MN} (\sig) d\sig^M d\sig^N = N^2 dz^2 + \chi_{\mu \nu} (d\sig^\mu + N^\mu dz) (d\sig^\nu + N^\nu dz)
\ee
between two hypersurfaces $\Sig_{\rm UV}$ and $\Sig_{\rm IR}$ at some constant-$z$ slices and
 whose respective intrinsic geometries are specified by $g_{ \mu \nu}$ and
$h_{\mu \nu}$, i.e.
\be \label{bdc0}
\chi_{\mu \nu} \bigr|_{\suv} 
=\bar g_{\mu \nu} (\sig^\lam), \qquad \chi_{\mu \nu} \bigr|_{\sir} 
= \bar h_{\mu \nu} (\sig^\lam) \ .
\ee
In~\eqref{path}, one should integrate over all values of $N$ and $N^{\mu}$ without any boundary conditions for them on $\suv$ and $\sir$. For this paper we will only be concerned with evaluating~\eqref{path} to leading order in the saddle point approximation, thus will not be careful about the precise definition of integration measure,  ghosts, and Jacobian factors for changes of variables. We will comment on these issues in Sec.~\ref{sec:diss}.

Later we will take $\suv$ to the boundary of an asymptotic AdS spacetime and $\sir$ to an event horizon.
We will for now keep them arbitrary for notational convenience.  We will also for now keep the gravitational action $S[G]$ general assuming only that it is diffeomorphism invariant and that the boundary conditions~\eqref{bdc0} give rise to a
well defined variational problem. The variation of the action then has the form
\be \label{var0}
\de S =  \ha \int d^{d+1} \sig \, \sqrt{-G} \, E^{MN} \de G_{MN}
\ee
without any boundary term. 
The equations of motion are thus
\be \label{eom0}
E^{MN}  = 0 \
\ee
while diffeomorphism invariance of the action $S [G]$ also leads to the Bianchi identities
\be \label{bianchi}
\nab_M E^{M N} = 0 \ .
\ee

Given the diffeomorphism invariance of the bulk action $S[G]$, $\Psi [\bar h, \bar g]$ is invariant under independent coordinate transformations ${\rm Diff}_{\rm IR} \times {\rm Diff}_{\rm UV}$ of the two hypersurfaces, i.e.
\be \label{hnem}
\Psi [\bar h, \bar g] = \Psi [\Lam_{\rm IR}^t \bar h \Lam_{\rm IR}, \Lam_{\rm UV}^t \bar g \Lam_{\rm UV}]
\ee
where $\Lam_{\rm IR}$ and $\Lam_{\rm UV}$ denote independent coordinate transformation matrices on
$\sir$ and $\suv$.



Now consider transforming the metric~\eqref{met0} to the Gaussian normal coordinates (GNC) $\xi^A (\sig) = (u, x^a)$ in  terms of which
\be \label{gnc0}
ds^2 =  du^2 + \ga_{ab} (u,x) dx^a dx^b
\equiv \tilde G_{AB} d\xi^A d\xi^B \ .
\ee
Here we choose Gaussian normal coordinates for later convenience.
The subsequent discussion applies with little changes to any set of  ``gauge fixed'' coordinates.
The metric components $G_{MN}$ can be expressed in terms of  $(\ga_{ab}, \xi^A)$ as
\be \label{irpr}
G_{MN} = \tilde G_{AB} \p_M \xi^A \p_N \xi^B =
\p_M u \p_N u + \ga_{ab} \p_M x^a \p_N x^b  \ .
\ee
In choosing the Gaussian normal coordinates we have the freedom
of fixing the values of $u$ and $x^a$ at one end. For our later purpose it is convenient to choose
a hybrid fixing
\be \label{bd1}
u \bigr|_{\suv} = u_0 = {\rm const}, \qquad x^a \bigr|_{\sir} = \sig^\mu \de^a_\mu.
\ee
 The values of $u$ at $\sir$ and $x^a$ at $\suv$ are then determined dynamically, which we will parameterize as
\be  \label{fixcor1}
u \bigr|_{\sir} =  \tilde \tau (\sig^\mu), \qquad x^a \bigr|_{\suv} = X^a (\sig^\mu)  \ .
\ee
In terms of the foliation of~\eqref{gnc0},  $\suv$ and $\sir$ can thus be written as
\be \label{irsur}
\suv: u = u_0, \qquad \sir: u = \tau (x^a) = \tilde \tau (X^{-1} (x^a))
\ee
and the boundary conditions~\eqref{bdc0} now become
\be \label{ubcd}
\ga_{ab} \big|_{u=\tau (x^a)} = h_{ab} , \qquad
\ga_{ab} \big|_{u=u_0} =  g_{ab} 
 \ee
with
\be      \label{bewbd}
h_{ab} = \bar h_{ab} - \p_a \tau \p_b \tau , \quad
g_{ab}  (X) = \bar g_{\mu \nu} (\sig) {\p \sig^\mu \ov \p X^a} {\p \sig^\nu \ov \p X^b}
= \le(J^{-1 t} \bar g J^{-1} \ri)_{ab}, \quad J^a{_\mu} \equiv {\p X^a \ov \p \sig^\mu} \ . 
\ee

Note that $\xi^A = (u (\sig^M), x^a (\sig^M))$ are dynamical variables and
in going from~\eqref{met0} to~\eqref{gnc0}, we have essentially traded $G_{MN} = (N, N_\mu, \chi_{\mu \nu})$ for
$(u, x^a, \ga_{ab})$.
The path integral can now be written as
\be \label{newpath}
\Psi [\bar h, \bar g] = \int Dx^a \int D u  \int_{ g}^{h} D \ga_{ab}  \, e^{{i } S [u, x^a, \ga_{ab} ]}
\ee
where $\ga$ is required to satisfy the boundary conditions~\eqref{ubcd}.
The coordinate invariance of the action implies that the action is independent of the
bulk fluctuations of $u$ and $x^a$. Thus the path integrals over $x^a$ and $u$ reduce to those over the boundary fluctuations~\eqref{fixcor1}
\be \label{pathb}
\Psi [\bar h, \bar g] = \int DX^a  \int D \tau \int_{ g}^{h} D \ga_{ab}  \, e^{{i} S [ \tau, \ga] }   \ .
\ee
In the above integrals $X^a$ always appears with $\bar g$ through the induced metric $g$ defined in~\eqref{bewbd}. In addition to appearing in the IR boundary condition $h_{ab}$ for $\ga_{ab}$ integrals, $\tau$ also appears in the action $S$ explicitly as the IR integration limit and boundary terms (which we will specify below).

$X^a$ and $\tau$ can be considered as the ``Wilson line'' degrees of freedom associated with $N^\mu$ and $N$.
Physically $X^a (\sig^\mu)$ describes the relative embedding between the coordinates $x^a$ on $\sir$ and
the coordinates $\sig^\mu$ on $\suv$, while $\tau (x)$ describes the proper distance between
$\suv$ and $\sir$.

The path integrals~\eqref{pathb} will be evaluated in stages: we first integrate over all possible $\ga_{ab}$ with a fixed relative embedding $X^a $ and proper distance $\tau$ to find
\be \label{path1}
e^{i I_{1} [\tau, h, g]} = \int_{ g}^{h} D \ga_{ab}  \, e^{{i} S [ \tau, \ga] }
\ee
and then integrate over $\tau$ (i.e. all possible proper distances)
\be \label{path2}
e^{i I_{\rm UV} [\bar h, g]} = \int D \tau \; e^{i I_{1} [\tau, h, g]}   \ .
\ee
In the path integrals~\eqref{path1}, with a finite $\tau$, $u$ direction is essentially compact and $\ga_{ab}$
can be consistently integrated out to yield a {\it local} action $I_1 [\tau, h, g]$, which can be expanded in the number of boundary derivatives of $\tau, h$ and $g$, assuming they are slowly varying functions of boundary coordinates.
In the boundary theory language $\ga_{ab}$ should thus correspond to modes with a mass gap.
$\tau$ depends only on boundary coordinates, and does not contain derivatives at leading order, and thus can also
be consistently integrated out. By definition $X^a$ always come with boundary derivatives as in~\eqref{bewbd}, i.e. they correspond to gapless modes, and thus should be kept in the low energy theory. Integrating them out will lead to nonlocal expressions.

Let us briefly consider the symmetries of $I_1 [\tau, g, h]$. It is invariant under $u$-independent diffeomorphisms of $x^a \to x'^a (x) $ under which
$g, h$ transform simultaneously as tensors.
These are large diffeomorphisms as they change the asymptotic behavior of AdS.
 $I_1$ is also invariant under diffeomorphisms
of $\sig^\mu$ as it contains $X^a$ and $\bar g$ only through $g$, which is invariant due to the canceling of transformations in $\bar g$ and $X^a$,
\be\label{now1}
\bar g_{\mu \nu} (\sig) \to \bar g_{\mu \nu}' (\sig) = {\p \sig'^{\lam} \ov \p \sig^\mu} \bar g_{\lam \rho} (\sig' (\sig))
 {\p \sig'^{\rho} \ov \p \sig^\nu} , \qquad X^a (\sig) \to X'^a (\sig) = X^a (\sig' (\sig)) \ .
\ee
This implies that
\be
\sqrt{-\bar g} \bar \nab_\nu \left( {1 \ov \sqrt{- \bar g} }
 {\delta  I_1  \ov \delta \bar g_{\mu \nu}}   \right) = {\de I_1 \ov \de X^a} \p^\mu X^a
 \ee
 where $\bar \nab$ denotes the covariant derivative associated with $\bar g$.
 Identifying $ {1 \ov \sqrt{- \bar g} }
 {\delta  I_1  \ov \delta \bar g_{\mu \nu}} $ as the boundary stress tensor, we then see that the $X^a$ equations of motion are equivalent to the conservation of the boundary stress tensor.
Parallel statements can be made about $I_{\rm UV}$ which comes from integrating out $\tau$.

Finally to conclude this subsection, let us be more explicit about the UV boundary condition.
In an asymptotic AdS spacetime with $\suv$ at a cutoff surface near the boundary, $\ga_{ab}$ in~\eqref{gnc0} should have the behavior 
\be
\lim_{u \to -\infty} \ga_{ab} (u, x^a) = e^{-{2 u\ov L} }  \le(\ga^{(0)}_{ab} (x^a) + O(e^{2 u \ov L}) \ri)
\ee
with $L$ the AdS radius and $\ga^{(0)}_{ab} (x^a)$ finite. Thus we should replace the boundary condition~\eqref{ubcd} at $u=u_0$ by
\be \label{adsbd}
\lim_{u_0 \to -\infty} \ga_{ab} (u_0, x^a) = e^{-{2 u_0 \ov L}} \le(g_{ab} (x) + O(e^{2u_0 \ov L})  \ri) ,
\ee
where
\be \label{gdef}
g_{ab} = \le(J^{-1 t} \bar g J^{-1} \ri)_{ab}
\ee
 and $\bar g_{\mu \nu} (\sig)$ is the background metric for the boundary theory.

\subsection{Saddle point evaluation} \label{sec:sadd}

Now consider evaluating the path integrals~\eqref{pathb}--\eqref{path2} using the saddle point approximation.
To elucidate the structure of equations of motion for $\tau$ and $X^a$, we consider~\eqref{var0}
now with $G_{MN}$ considered as a function of $\ga_{ab}$ and $\xi^A$ via~\eqref{irpr}.
Under variations of $\ga_{ab}$, we have
\be
\de G_{MN} (x) = \de \ga_{ab}   \p_M x^a \p_N x^b
\ee
which then~\eqref{var0} implies the equations of motion
\be \label{dyn0}
E^{MN}  \p_M x^a \p_N x^b  = 0 , \quad \Rightarrow \quad E^{ab} = 0
\ee
where $E^{ab} =0$ is the $ab$-component of the equations of motion in coordinates~\eqref{gnc0}. Below we will refer to these
equations as ``dynamical equations.''

Under variations of $\xi^A$, we have
\be
\de G_{MN} = {\p \tilde G_{AB} \ov \p \xi^C} \p_M \xi^A \p_N \xi^B \de \xi^C + 2 \tilde G_{AC} \p_M \xi^A \p_N \de \xi^C  \ .
\ee
The bulk part of~\eqref{var0} then leads to the Bianchi identities in coordinates~\eqref{gnc0}, which is as it should be since $\xi^A (\sig)$ is a coordinate transformation. But now there are boundary terms remaining
\be
\de S = \int d^d x \, \sqrt{-H} E_A{^B} {\p z \ov \p \xi^B} \de \xi^A \biggr|_{\sir} -  \int d^d x \, \sqrt{-H} E_A{^B} {\p z \ov \p \xi^B} \de \xi^A \biggr|_{\suv}
\ee
which upon using~\eqref{bd1}--\eqref{irsur} implies that
\be \label{cons0}
\sqrt{-H}  E_u{^B} {\p z \ov \p \xi^B} \biggr|_{\sir} = 0 \quad \Rightarrow \quad
\le(E_u{^u} - E_u{^a} {\p \tau \ov \p x^a} \ri) \biggr|_{\sir} = 0
\ee
and
\be \label{cons1}
\sqrt{-H}  E_a{^B} {\p z \ov \p \xi^B} \biggr|_{\suv} = 0   \quad \Rightarrow \quad  E_a{^u} \bigr|_{\suv} = 0  \
\ee
with~\eqref{cons0} corresponding to the equation of motion from varying $\tau$ while~\eqref{cons1}
 corresponds to those from varying $X^a$. In deriving the second equations in both~\eqref{cons0} and~\eqref{cons1}
we have assumed that $\sqrt{-H}$ and ${\p z \ov \p u}$ at $\sir$ and $\suv$ are nonzero.
It can be readily checked that~\eqref{dyn0} and~\eqref{cons0}--\eqref{cons1} are equivalent to~\eqref{eom0}, and that
the Bianchi identity ensures that~\eqref{cons0} and~\eqref{cons1} are satisfied everywhere once they are imposed
at $\sir$ and $\suv$, respectively. Following standard convention, below we will refer to~\eqref{cons0} as the Hamiltonian constraint and~\eqref{cons1} as the momentum constraints.

Now recall from the general
results on the holographic stress tensor~\cite{Balasubramanian:1999re} that the momentum constraints~\eqref{cons1} in fact correspond to the conservation of the boundary stress tensor
\be \label{cond0}
\nab_a T^{ab} = 0
\ee
where $\nab_a$ is the covariant derivative associated with $g_{ab}$ and $T^{ab}$ is the  stress tensor
for the boundary theory with background metric $g_{ab}$ in the state described by~\eqref{gnc0}.
Since $\bar g_{\mu \nu}$ and $g_{ab}$ are related by a coordinate transformation equation~\eqref{cond0} is  equivalent to
\be \label{cond1}
\bar \nab_\mu \bar T^{\mu \nu} = 0
\ee
where $\bar \nab_\mu$ is the covariant derivative associated with $\bar g_{\mu \nu}$ and $\bar T^{\mu \nu}$ is
the stress tensor for the boundary theory with background metric $\bar g_{\mu \nu}$. This gives an alternative way to see
that $X^a$ equations of motion are equivalent to conservation of the boundary stress tensor.

At the level of saddle point approximation, $I_1$ as defined in~\eqref{path1} is obtained by solving~\eqref{dyn0} for $\ga_{ab}$, and $I_{\rm UV}$ in~\eqref{path2} by solving~\eqref{cons0} for $\tau (x^a)$.
In other words, $I_{\rm UV}$ is computed by evaluating the gravity action with dynamical equations and the Hamiltonian constraint imposed, but not the momentum constraints.

\subsection{Einstein gravity}\label{sec:ein}

We now specialize to Einstein gravity, in which the gravity action $S[G]$ in~\eqref{path} can be written in
 the Gaussian normal coordinates~\eqref{gnc0} as
\bea
S [\tau, \ga] &=&  \int d^{d+1} \xi\, \sqrt{-\ga} (R - 2 \Lam) - \int_{\suv} \sqrt{-\ga}\, 2 K + \int_{\sir} \sqrt{-\ga} \, 2 K  + S_{\rm ct}
\cr
& = &
 \int d^{d+1} \xi \,
 \sqrt{-\ga}  \left[ ^{(d)} R [\ga]-  \le( K_{ab} K^{ab} - K^2 \ri)- 2 \Lam \right]   + S_{\rm IR}  + S_{\rm ct} 
 \label{ehact}
 \eea
where
\be \label{extde}
K_{ab} = \ha \p_u \ga_{ab} \equiv \ha \ga_{ab}', \qquad K^a{_b} = \ha (\ga^{-1} \ga')^a{_b}, \qquad K = K^a{_a} =  \ha \ga^{ab} \ga_{ab}'
\ee
are extrinsic curvatures for a constant-$u$ hypersurface, and $S_{\rm ct}$ is the standard AdS counterterm action at the $\suv$ \cite{Balasubramanian:1999re}
\be
S_{\rm ct} =  \int_{u = u_0 \to -\infty} d^d x \sqrt{-\ga} \, \le({2 (1-d) \ov L}+ {L \ov d-2}  {^{(d)}}\!R [\ga] + \cdots \ri)   \ .
\ee
$S_{\rm IR}$ is a boundary action at $\sir$ which arises from the fact that $\sir$, given by $u = \tau (x^a)$, is not compatible with the foliation of constant-$u$ hypersurfaces, and can be written as (see Appendix~\ref{app:bd} for a derivation)
\be
S_{\rm IR} = 2 \int_{\sir} d^d x \sqrt{-\bar h} \, {1 \ov \fn} \le( \le( \bar h^{ab} (\bar\p \tau)^2  -  \bar\p^a \tau \bar \p^b \tau  \ri) K_{ab}
-  \bar D^2 \tau \ri)
\ee
with
\be
\fn  = \sqrt{1 - \bar\p^a \tau \p_a \tau}
\ee
where the indices are raised and lowered by the intrinsic metric $\bar h_{ab}$ on $\Sigma_{\rm IR}$  and $\bar D$ is the covariant derivative associated with $\bar h_{ab}$.

For convenience below we will use $\sK$ to denote the matrix $K^a{_b}$ and thus $K = \Tr \sK$.
Various components of the Einstein equations in Gaussian normal coordinates~\eqref{gnc0}
can then be written as
\begin{gather} \label{dynaG}
- E^{a}_{b} = \sK'-\mbox{Tr}\sK'+\sK\mbox{Tr}\sK-\mbox{Ric}^{(d)} [\ga]-\frac{1}{2}\left(\mbox{Tr}\sK^{2}+\mbox{Tr}^{2}\sK\right)+\frac{1}{2}R^{(d)} [\ga] - \Lam = 0 \\
 \label{hamG}
- E^{uu} = \frac{1}{2}\mbox{Tr} \sK^{2}-\frac{1}{2}\mbox{Tr}^{2}\sK+\frac{1}{2}R^{(d)} [\ga] - \Lambda =0  \\
\label{momG}
- E^{u}{_a} = \sD_a K - \sD_b K^{b}{_a}=0
\end{gather}
with $\sD_a$ the covariant derivative associated with $\ga_{ab}$.
 As discussed in Sec.~\ref{sec:sadd}, in order to not impose conservation of the stress tensor, i.e. leave hydrodynamical modes off-shell, at the saddle point level 
 we should {\it not impose} the momentum constraint~\eqref{momG}. We only need to solve the dynamical equations~\eqref{dynaG} for $\ga_{ab}$  and a combination of~\eqref{hamG}--\eqref{momG}  at $\sir$ for $\tau$ (see~\eqref{cons0}).

From now on we will set the AdS radius $L =1$.

\section{Action for an ideal fluid}\label{thirdsec}

In this section we first evaluate explicitly $I_{\rm UV} [X^a, \bar g, \bar h]$ defined in~\eqref{uvint1}
at zeroth order in the derivative expansion, assuming that $X^a, \bar g, \bar h$  are slowly varying functions. We then show that pushing $\bar h$ to a horizon hypersurface and
requiring it to be non-dissipative, we obtain the ideal fluid action of~\cite{Dubovsky:2005xd,Dubovsky:2011sj}.



\subsection{Solving the dynamical equations}

We will perform the $\ga$ integrals~\eqref{path1} using the saddle point approximation, i.e. it boils down to solving the dynamical Einstein equations~\eqref{dyn0} at zeroth order in boundary derivatives.
At this order we can neglect boundary derivatives of $\tau (x)$, $J^a{_\mu}$, and $\ga_{ab}$. The boundary conditions for $\ga_{ab}$ become
\be \label{bdss0}
 \ga (u = \tau) = h = \bar h, \qquad \ga (u = u_0 \to -\infty) = e^{-{2 u_0 }} g  , \qquad
 g_{ab} = \le(J^{-1 t} \bar g J^{-1} \ri)_{ab} \ .
\ee
For notational simplicity here and below we will often use $\ga$ and $g$ to denote the whole matrix $\ga_{ab}$ and $g_{ab}$. Hopefully the context is sufficiently clear that they will not be confused with their respective determinants.

Explicit expressions for the Einstein equations in Gaussian normal coordinates~\eqref{gnc0} are given
in Sec~\ref{sec:ein}. At zeroth order in boundary derivatives, the dynamical part of the Einstein equations~\eqref{dyn0} (more explicitly~\eqref{dynaG}) becomes
\be
\sK'-\mbox{Tr}\sK'+\sK\mbox{Tr}\sK - \frac{1}{2}\left(\mbox{Tr}\sK^{2}+\mbox{Tr}^{2}\sK\right) - \Lam = 0
\ee
which can be rewritten as 
\bea
\label{trace}
&& {d-1 \ov d} \left(K'+ \ha K^2\right)=-\Lambda-\ha \Tr \KK^2, \\
\label{traceless}
&& \KK' + K  \KK=0
\eea
 where $\KK$ is the traceless part of $\sK$
 \be \label{neoeo}
 \sK = \KK + {K \ov d} \mathbf{1}, \qquad \Tr \KK =0  \ .
 \ee
From~\eqref{traceless}
 \be \label{e1}
 \ha (\Tr \KK^2)' = - K \Tr \KK^2  \ .
 \ee
Taking derivative on both sides of~\eqref{trace} leads to
\be  \label{tt1}
 {d-1 \ov d} \left(K''+  K K' \right)=K \Tr \KK^2 \ .
 \ee
 Eliminating $\Tr \KK^2$ between~\eqref{e1}--\eqref{tt1} and using~\eqref{trace} we then find an equation for
 $K$
 \be
 K'' + 3 K' K + K (K^2 - d^2) = 0
  \ee
which is solved by
\be \label{hhe}
K = d { \al_1 \al_2  e^{2 du} -1  \ov (1+ \al_1 e^{du}) (1 + \al_2 e^{du})}
\ee
where $\al_{1,2}$ are some constants.
Inserting~\eqref{hhe} into~\eqref{traceless} we find
\be \label{bb1}
\KK=\frac{b_0 (\al_1 - \al_2) e^{du}}{ (1+ \al_1 e^{du}) (1 + \al_2 e^{du})}
\ee
with $b_0$ a constant traceless matrix. Plugging~\eqref{bb1} into~\eqref{tt1} we find that
$b_0$ has to satisfy
\be \label{co3}
\Tr b_0^2 = d (d-1) \ .
\ee
Combining~\eqref{hhe} and~\eqref{bb1} we then obtain
\be \label{sk0}
\sK = {1 \ov (1+ \al_1 e^{du}) (1 + \al_2 e^{du})} \le(b_0  (\al_1 - \al_2)  e^{du} + (\al_1 \al_2  e^{2 du}-1) \ri) \ .
\ee

Now integrating~\eqref{extde} and imposing the boundary condition at UV (i.e $u=u_0 \to -\infty$) we find that
\be\label{onrl}
\ga = g e^{-2 u} \le( (1 + \al_1 e^{du}) (1 + \al_2 e^{du}) \ri)^{2 \ov d} e^{2k (u) b_0}
\ee
with
\be
k (u) = {1 \ov d}   \log \le({1 + \al_1 e^{du} \ov
1 + \al_2 e^{du}} \ri)
\ee
 where here and below we will always take $\al_1 > \al_2$.

Introducing
\be
H \equiv g^{-1} \bar h, \qquad \al_{1c} \equiv \al_1 e^{d \tau} , \qquad \al_{2c} \equiv \al_2 e^{d \tau}
\ee
from the IR boundary condition $\ga (u=\tau) = \bar h$ we then find
\be \label{igne}
\sqrt{\det H} =e^{- d \tau} (1+ \al_{1c})(1+\al_{2c}) , \quad b_0 = {1 \ov 2 k (\tau)} \log \hat H , \quad
\Tr (\log \hat H)^2 =  {4(d-1) \ov d} \log^2 { 1+ \al_{1c} \ov 1+ \al_{2c}}
\ee
where $\hat H$ denotes the unit determinant part of $H$ and the last equation of~\eqref{igne} follows from~\eqref{co3}. Requiring the metric $\ga_{ab}$ to be regular and non-degenerate  between $u =-\infty$ and $\tau$, we need
\be
1+ \al_{1c} > 0, \qquad 1 + \al_{2c} > 0 \ .
\ee

Given $H$ and $\tau$, we can use the first and last equations of~\eqref{igne} to determine $\al_{1,2}$ and then the second equation to find $b_0$. Note at this stage  $\tau$ is not constrained by $H$ and thus can be chosen independent of $\bar h$.
More explicitly,
\be \label{a12c}
\al_{1c} = (\det H)^{1 \ov 4} e^{z_c \ov 2} e^{d\tau \ov 2} - 1, \qquad
\al_{2c} = (\det H)^{1 \ov 4} e^{-{z_c \ov 2}} e^{d\tau \ov 2} - 1
\ee
with
\be\label{zcdef}
z_c \equiv  \le({d \ov 4 (d-1)} \Tr (\log \hat H)^2 \ri)^\ha \ .
\ee

\subsection{Effective action for $\tau$ and $X^a$}

At zeroth order in boundary derivatives, the Einstein action~\eqref{ehact} becomes
\be \label{ehact1}
S [\tau, \ga] =   \int d^{d+1} \xi \,
 \sqrt{-\ga}  \left[ -  \le(\Tr \sK^2 - K^2 \ri)- 2 \Lam \right]  + S_{\rm ct} 
 \ee
and substituting~\eqref{sk0} into~\eqref{ehact1} we have
\be
I_1 [\tau, \bar h, g] =  2 (d-1) \int d^d x \sqrt{-g} \le[-e^{- d \tau} + e^{-d \Lam} + \al_1 \al_2  e^{d \tau} \ri] + S_{\rm ct}
\ee
with the counterterm action given by
\be
S_{\rm ct} =- 2 (d-1) \int_{\Lam \to -\infty} d^d x \sqrt{-\ga} =
- 2 (d-1) \int d^d x \sqrt{-g} \, \le(e^{-d \Lam} + \al_1 + \al_2 \ri)  \ .
\ee
We then find that
\be
I_1 [\tau,  \bar h, g] =  - 2 (d-1) \int d^d x \sqrt{-g} \, \sL_1 (H, \tau)
\ee
with
\bea
\sL_1 (H, \tau) &=& e^{- d \tau} + \al_1 + \al_2  - \al_1 \al_2  e^{d \tau}\cr
& = & - \sqrt{\det H} + 4 e^{- \ha d \tau} (\det H)^{1 \ov 4} \cosh {z_c \ov 2}- 2 e^{-d \tau}
\eea
where in the second line we have expressed the integration constants $\al_{1,2}$ in terms of boundary conditions via~\eqref{a12c}. We notice that at zeroth order, $I_1$ depends on $\bar h$ and $g$ only through the combination $H = g^{-1} \bar h$. This follows from that $I_1$ must be invariant under the diffeomorphisms of $x^a$ for which $\bar h$ and $g$
transform simultaneously, as noted in the paragraph before~\eqref{now1}. At zeroth order $\Tr H^n$ for $n=1,2,\dots,d$ are the only independent invariants.

$\tau$ can now be integrated out by extremizing $I_1$ which gives
\be \label{tauso}
e^{- d\tau_0} = \sqrt{\det H} \cosh^2 {z_c \ov 2}
\ee
and thus
\be
\sL_{\rm UV}  = \sL_1 (H, \tau_0) = \sqrt{\det H} \cosh z_c  \ .
\ee
Collecting everything together we thus find that\footnote{The overall minus sign has to do with our choice of orientation of bulk manifold.}
\be \label{preact}
I_{\rm UV} [X^a; \bar h, \bar g] =    -  2(d-1) \int d^d x \sqrt{-g} \, \sqrt{\det H} \cosh \le({d \ov 4 (d-1)} \Tr (\log \hat H)^2 \ri)^\ha \ .
\ee
One can readily check that the same result is obtained by solving instead the Hamiltonian constraint~\eqref{cons0} at zeroth order. Also note that with $\tau = \tau_0$ given by~\eqref{tauso}, equation~\eqref{a12c} becomes
\be \label{oeen1}
\al_{1c} = - \al_{2c} = \tanh {z_c \ov 2} \ .
\ee

\subsection{Horizon limit}

Equation~\eqref{oeen1} implies that after integrating out $\tau$, $\al_2 = - \al_1 $ is negative. We will now simply rename $\al_1$ as $\al$. It is convenient to introduce
\be \label{ooem}
  \al \equiv e^{- d u_h}
 \ee
 then from~\eqref{oeen1}
 \be
u_h = \tau - {1 \ov d} \log \tanh {z_c \ov 2} > \tau \ .
\ee
 Now if we extrapolate the solution~\eqref{onrl} beyond $u =\tau$ all the way to $u_h$, then $\sqrt{-\det \ga}$ develops a simple zero at $u= u_h$, which we will refer to as a ``horizon.'' Since the ``horizon'' lies outside the region where our Dirichlet problem is defined, $\ga_{ab}$ does not have to be regular there, so this does not have to be the horizon in the standard sense. Now let us consider a sequence of $\bar h$ whose time-like eigenvalue approaches zero. Equivalently,
an eigenvalue of $H$ (which we will denote as $h_0$) goes to zero. At $h_0=0$, $\bar h$ describes
a null hypersurface and $\sir$ becomes a horizon for the metric between $\sir$ and $\suv$.
We thus define the $h_0 \to 0 $ limit as the hydrodynamic limit.

In this limit, we have
\be
\det H \to h_0 \sdet H , \qquad z_c \to - \ha \log h_0 + {1 \ov 2 (d-1)} \log \sdet H   \
\ee
where $\sdet H$ denotes the non-vanishing subdeterminant of $H$ and can be written as
\be \label{ueu}
\sdet H = p_{d-1} \le(\Tr H , \Tr H^2 , \cdots \ri) \
\ee
where $p_{d-1}$ is the standard polynomial which expresses the determinant of a non-singular $(d-1) \times (d-1)$ matrix in terms of its trace monomials.
From~\eqref{tauso} and~\eqref{ooem} we thus find
\be\label{uhs1}
e^{- d \tau_0} \to {1 \ov 4} (\sdet H)^{d \ov 2 (d-1)} , \qquad u_h - \tau \to 0
\ee
and the action~\eqref{preact} becomes
\bea
I_{\rm UV} &=&   -  (d-1) \int d^d x \sqrt{-g} \, (\sdet H)^{d \ov 2 (d-1)} \\
&=& - (d-1) \int d^d \sig  \sqrt{-\bar g}  \, (\sdet H)^{d \ov 2 (d-1)} \ .
 \label{hydro0}
\eea

\subsection{Entropy current}

Here we discuss the geometric meaning of $\sdet H$ and~\eqref{hydro0}.
Denote the null eigenvector of $\bar h$ by $\ell^a$, which give rises to a congruence of null geodesics which generate the null hypersurface.
We can then choose a set of coordinates  $(v, \xi^\rmi)$ on $\sir$ with $v$ the parameter along the null geodesics generated by $\ell^a$ and  $\xi^\rmi$ remaining constant along geodesics. In this basis, we then write the metric on $\sir$ as
 \be \label{irmep}
ds^2_{\sir} = \bar h_{ab} dx^a dx^b = \sig_{\rmi \rmj} (v, \xi) d\xi^\rmi d \xi^\rmj , \quad \bar h_{ab} = \sig_{\rmi \rmj} {\p \xi^\rmi \ov \p x^a} {\p \xi^\rmj \ov \p x^b} , \quad \rmi =1,2, \cdots n, \quad n \equiv d-1
  \ .
 \ee
It then follows that
\be
\sdet H = \det (\al^{\rmi\rmj} \sig_{\rmj \rmk}) = \det \sig \det \al^{-1}
\ee
where $\al^{-1}$ is defined as
\be \label{aldef}
(\al^{-1})^{\rmi \rmj} \equiv \al^{\rmi \rmj} = \bar g^{\mu \nu} E^\rmi{_\mu} E^\rmj{_\nu}, \quad
E^\rmi{_\mu} \equiv  {\p \xi^\rmi \ov \p x^a} J^a{_\mu} d\sig^\mu = {\p \xi^\rmi \ov \p \sig^\mu} \ .
\ee
 We thus find that $\sqrt{\sdet H}$ can be written as horizon area density $\sqrt{\sig}$ normalized by the ``area density'' of the pull back of boundary metric $\bar g$ to $\sir$.

The physical meaning of $\sqrt{\sdet H}$ can be better elucidated if instead we pull back all quantities to the boundary.
We now show that it can be interpreted as a definition of (non-equilibrium) entropy density of the boundary system.\footnote{The construction below is similar to the construction of entropy current in fluid/gravity~\cite{Bhattacharyya:2008xc,Hubeny:2011hd}.}
For this purpose, consider the area form on the $\sir$ which can be written as
\be \label{aref}
a = \sqrt{\sig} \, d\xi^1 \wedge d \xi^2 \wedge \cdots \wedge d\xi^n \ .
\ee
Note that the horizon area $\sqrt{\sig}$ has no physical meaning itself in the boundary theory as its definition depends on a
choice of local basis. It does become a physically meaningful quantity when we pull it back to the boundary via the relative embedding map $J^a{_\mu}$ introduced in~\eqref{bewbd}. More explicitly,
\be \label{area1}
 a = \sqrt{\sig} E^1 \wedge E^2 \wedge \cdots \wedge E^n, \qquad  E^\rmi =
 E^\rmi{_\mu} d\sig^\mu \ .
\ee
From~\eqref{area1} we can define a current which is the Hodge dual of $a$ on the boundary
\be \label{encu0}
j^\mu = \ep^{\mu\nu_1 \cdots \nu_n}  a_{\nu_1 \cdots \nu_n} = {1 \ov n!} \ep^{\mu\nu_1 \cdots \nu_n} \ep_{\rmi_1 \cdots \rmi_n}  E^{\rmi_1}{_{\nu_1}} \cdots E^{\rmi_n}{_{\nu_n}}
\ee
where $\ep^{\mu\nu_1 \cdots \nu_n} $ is the full antisymmetric tensor for $\bar g$ and
$\ep_{\rmi_1 \cdots \rmi_n}$ is that for $\sig_{\rmi \rmj}$. Similarly, it is natural to pull back the null vector $\ell^a$ to the boundary, giving
\be
u^\mu = (J^{-1})^\mu{_a} \ell^a, \qquad \bar g_{\mu \nu} u^\mu u^\nu = -1 \
\ee
and we have chosen a convenient normalization for $u^\mu$. By construction, $j^\mu$  is parallel to $u^\mu$ and we can then write
\be\label{encu1}
j^\mu = s u^\mu, \qquad s^2 = - j^\mu j_\mu  \ .
\ee
From~\eqref{encu0}, we find that
\be \label{ebd}
s = \sqrt{\det (\al^{\rmi\rmj} \sig_{\rmj \rmk})}  = \sqrt{\sdet H} \ .
\ee
We will interpret $u^\mu$ as the velocity field of the boundary theory, $j^\mu$ (divided by $4 G_N$) as the entropy current, and $s$ (divided by $4 G_N$) as the local entropy density. All these quantities are independent of choice of local coordinates on $\sir$. We also stress that their definitions do not depend on the derivative expansion and thus should apply to all orders.

With this understanding the action~\eqref{hydro0} can be written as
\be
I_{\rm UV}  = - \int d^d \sig  \sqrt{-\bar g}  \, \ep (s)  \ .
\ee
where
\be \label{onun}
\ep (s) = (d-1) s^{d \ov d-1}
\ee
has precisely the scaling of the local energy density as a function of entropy density for a conformal theory.
From the perspective of evaluating the bulk action it can also be understood as follows: the bulk integration in~\eqref{ehact}
can be interpreted as giving the free energy while the Gibbons-Hawking term at the IR hypersurface becomes equal to entropy times temperature in the horizon limit. Their sum then gives the energy of the system.


\subsection{Hydrodynamical action and volume-preserving diffeomorphism}

We now impose that the system is non-dissipative, which amounts to requiring that the entropy current~(\ref{encu0})
is conserved
\be \label{nondih1}
\bar \nabla_\mu j^\mu=0.
\ee
where $\bar \nab$ is the covariant derivative for $\bar g$. The above equation can also be written equivalently
in various different ways in terms of horizon quantities. In terms of the horizon area form (\ref{aref}),
\be
da=0
\ee
or area density
\be \label{nondih}
\p_v \det \sig = 0 \quad \to \quad \det \sig = \det \sig (\xi) \ .
\ee
i.e. the horizon area is independent of the horizon ``time'' $v$.
Note that the form of the metric~\eqref{irmep}
is preserved with a $v$-independent coordinate transformation
\be \label{uenl}
\xi^\rmi \to \xi'^\rmi (\xi)
\ee
which can be used  to set
\be \label{oorn}
\det \sig = 1  \quad \to \quad \sdet H = \det \al^{-1} .
\ee
We thus  see that with non-dissipative boundary condition at zeroth order  the horizon metric {\it completely decouples}
in the hydrodynamical action, and we find
\be
I_{\rm UV}  = - (d-1) \int d^d \sig  \sqrt{-\bar g}  \, \le(\det \al^{-1} \ri)^{d \ov 2 (d-1)}, \qquad
(\alpha^{-1})^{\rmi \rmj}=\bar g^{\mu\nu}   \partial_\mu \xi^{\rmi}  \partial_\nu  \xi^{\rmj}
\ee
which is precisely that of~\cite{Dubovsky:2005xd,Dubovsky:2011sj} applied to a conformal theory.

 After fixing~\eqref{oorn}, there are still residual volume-preserving diffeomorphisms in $\xi^\rmi$, which played an important role in the formulation of~\cite{Dubovsky:2005xd,Dubovsky:2011sj}.  Here they arise out of a subgroup of horizon diffeomorphisms which leave ``gauge fixing condition''~\eqref{oorn}
and the coordinate frame~\eqref{irmep} invariant.
If the non-dissipative horizon condition~\eqref{nondih} can be consistently imposed to higher
orders in derivative expansion, we should expect the resulting higher order action to respect the volume-preserving diffeomorphisms. As we will see in Sec.~\ref{sec:2nd}, however, at second order we encounter divergences, which implies
 that~\eqref{nondih} can no longer be consistently imposed for unconstrained integrations of $X^a$.

\subsection{More on the off-shell gravity solution}


Here we elaborate a bit further on the off-shell gravity solution~\eqref{onrl}.
First let us collect various earlier expressions.
After integrating out $\tau$, with~\eqref{oeen1} and~\eqref{ooem} the off-shell metric~\eqref{onrl} can be written as
\be \label{offmetric}
\ga = g e^{-2 u} \le(1 - \al^2 e^{2du} \ri)^{2 \ov d} e^{{z (u) \ov z_c} \log \hat H}
\ee
where we have introduced
\be \label{uieop}
z (u) =  \log \le({1 + \al e^{du} \ov
1 - \al e^{du}} \ri) =  - {\log \tanh {d  \ov 2} (u_h -u)} 
= \tanh^{-1}  \al e^{du}   \ . 
\ee
Also recall that
\be\label{defD}
 e^{- d \tau} = \sqrt{\det H} \cosh^2 {z_c \ov 2} , \quad \al =e^{-d u_h} = \ha \sqrt{\det H} \sinh z_c , \quad
\De \equiv  d (u_h - \tau) = -  \log \tanh {z_c \ov 2} \ .
\ee

With a general regular $\bar h$, the solution when extrapolated to the ``horizon'' $u_h$ is singular.
This is perfectly okay as physically the behavior of the metric outside the region is of no concern to us.
We will now show that if we do require the extrapolated metric to be also regular at the ``horizon'' $u = u_h$, i.e. $u=u_h$ becomes a genuine horizon, then we recover the standard black brane solution. Of course this also implies that $\bar h$ has to take a very specific form.

We now impose a ``regularity'' condition: $\ga_{ab}$ has only one eigenvalue approaching zero
as the horizon is approached with the other eigenvalues finite. Near the horizon, $\de \equiv d(u_h - u) \to 0$ with
\be
z (u) \to - \log {\de \ov 2} \to +\infty \ .
\ee
Denoting the eigenvalues of $\log \hat H$ and $g^{-1} \ga $ as $\hat b_\mu$ and $\ga_\mu$ respectively, from~\eqref{offmetric}
we then have
\be
\ga_\mu \to  e^{- 2 u_h}  (2 \de)^{2 \ov d}  \le({\de \ov 2} \ri)^{-{\hat b_\mu \ov  z_c}} 
\ .
\ee
If we denote the time-like eigenvalue of $g^{-1} \ga$ by $\ga_0$ and the rest by $\ga_i$, the regularity condition amounts to that $\ga_0$ goes to zero while all $\ga_i$ finite.
Requiring $\ga_i$ to be finite implies that
\be \label{ebe}
\hat b_i =  {2 z_c \ov d} =  - {2 \ov d} \log  \tanh {\De \ov 2} \equiv b , \quad \hat b_0 = - {2 (d-1) \ov d} z_c = - (d-1) b
\ee
where the second equation follows from that  $\log \hat H$ is traceless. We thus find that the system has to be isotropic!

Denoting the time-like eigenvector vector of $\hat H$ as $\ell^a$, from~\eqref{ebe} we can write $\hat H$
as
 \be \label{onem}
\hat H^a{_b} = (d-1) b \ell^a\ell_b + b (\de^a{_b} + \ell^a \ell_b )
\ee
where we have defined
\be
\ell_a = g_{ab} \ell^b, \qquad \ell^a \ell_a = -1 \ .
\ee
Plugging~\eqref{onem} into~\eqref{offmetric} we then find that
\be
\ga_{ab} =  C \rho^{4 \ov d} \le( g_{ab} +\rho^{-2} \ell_a \ell_b \ri)
\ee
with
\be
 C = e^{-2 u_h} 2^{4 \ov d}  , \qquad \rho(u) = \cosh {d (u_h - u) \ov 2}  \ .
 \ee
This is precisely the black brane metric and $\ell^a$ is the null vector of the horizon hypersurface.

Consider an arbitrary basis of vectors $E_{\rmi a}$ which are orthogonal to $\ell^a$, we can expand
\begin{gather} \label{jne}
g_{ab} = - \ell_a \ell_b +\al^{\rmi \rmj}  E_{\rmi a}  E_{\rmj b} \\
\bar h_{ab} 
= - h_0 \ell_a \ell_b +  \sig^{\rmi \rmj}  E_{\rmi a}  E_{\rmj b}
\label{jne1}
\end{gather}
then equation~\eqref{onem} implies that
\be\label{regcond01}
 \sig^{\rmi \rmk} \al_{\rmk \rmj} = c \de_\rmi{^\rmj} \
\ee
where $\al_{\rmi \rmj}$ is the inverse of $\al^{\rmi \rmj}$ and $c$ is some constant. In other words, regularity condition fixes $\bar h$ in terms of $g$ up to
two constants $c$ and $h_0$. $u_h$ and $\tau$ can be expressed in terms of $c$ and $h_0$ as
\bea
e^{du_h}={4c^{1-{d\over 2}}\over c-h_0}, \qquad e^{d\tau}={4c^{1-{d\over 2}}\over (\sqrt{c }+\sqrt{h_0})^2} \ .
\eea

Given that $\ell^a$ is the null vector of the horizon and we can choose basis 
$E_{\rmi a}$ in~\eqref{jne} to be that in~\eqref{irmep} (with $\rmi$ index raise and lowered by $\al$) 
and then $\al^{\rmi \rmj}$ of~\eqref{jne} then coincides that in~\eqref{aldef}, and thus the same notations. Similarly in the horizon limit $h_0 \to 0$, and $\sig^{\rmi \rmj}$ 
of~\eqref{jne1} is related to $\sig_{\rmi \rmj}$ in~\eqref{irmep} by raising and lowering  using $\al$.

\section{Generalization to higher orders} \label{sec:2nd}

\newcommand{\tx}[1]{\text{#1}}

In this section we discuss computation of $I_{\rm UV}$ to higher orders in the derivative expansion.
We first briefly outline the general structure of higher order calculations and then mention some results at second order.

\subsection{Structure of derivative expansions to general orders} \label{sec:high}

Assuming $\bar h, \bar g$ and $X^a$ are slowly varying functions of boundary spacetime variables, we can 
expand $\ga_{ab}$, the extrinsic curvature $\sK$, and $\tau$ in the number of boundary derivatives, i.e. 
\be \label{exnd}
\ga = \ga_0 + \ga_2 + \cdots, \qquad \sK = \sK_0 + \sK_2 + \cdots , \qquad \tau = \tau_0 + \tau_2 + \cdots  \ .
\ee
where $\ga_0, \sK_0, \tau_0$ (which we already worked out) contain zero boundary derivatives of $\bar g, \bar h, J^a{_\mu} = \p_\mu X^a$, whereas $\ga_2, \sK_2, \tau_2$ contain two boundary derivatives, and so on.  One can readily see that there is no first order contribution, as the equations for the saddle point (\ref{dynaG}) and (\ref{hamG}) do not have first order terms,
and neither does the action~\eqref{ehact}. The final hydrodynamical action~\eqref{enne} will receive first order contributions as
the IR contribution $\Psi_{\rm IR}$ will contain first order terms, which will communicate via matching to $\Psi_{\rm UV}$ at the stretched horizons through equations for $\bar h_1$ and $\bar h_2$. 

Let us first look at the dynamical equations~\eqref{dynaG} which under decomposition~\eqref{neoeo} can be written as 
\bea \label{K1}
&& {d-1 \ov d} \left(K'+ \ha K^2\right)=-\Lambda-\ha \Tr \KK^2 + {d-2 \ov 2d} \dR  , \\
&& \KK' + K  \KK= \sR- {1 \ov d} \dR \ 
\label{K2}
\eea
where $\sR$ denotes the matrix of mixed-index Ricci tensor $\dR^a{_b}$. Taking the $u$ derivative on~\eqref{K1} and using~\eqref{K1}--\eqref{K2} we find that
\be \label{K3}
K'' + 3 K' K + (K^2 -d^2 ) K = \dR K + {d-2 \ov 2 (d-1)} \dR' - {d \ov d-1} \Tr \sR \sK \ . 
\ee
Plugging~\eqref{exnd} into these equations we find at $n$-th order
\bea \label{K4}
&& K''_{n} + 3 K_0 K_n' + (3K_0' + 3 K_0^2 -d^2) K_n = S_n\\
&& \KK_n' + K_0 \KK_n + K_n \KK_0 = P_n
\label{K5}
\eea
where sources $S_n$ and $P_n$ contain only quantities which are already solved at lower orders. 
Note that $P_n$ is a traceless matrix. Parallel to earlier zeroth order manipulations, the integration constants in $\KK_n$ will need to satisfy a constraint from~\eqref{K1}
\be \label{K6}
{d-1 \ov d} \left(K'_n+ K_0 K_n \right) + \Tr \KK_0 \KK_n = B_n 
\ee
where $B_n$ again contains only quantities solved at lower orders. Thus once we have solved the nonlinear equations at the zeroth order, higher order corrections can be obtained by solving linear equations. In particular, at each order the 
homogeneous part of the linear equations are identical with only difference being the sources. This aspect is very similar to the structure of equations in the fluid/gravity approach~\cite{Bhattacharyya:2008jc}. For completeness we give explicit expressions of various sources $S_n, P_n, B_n$ in Appendix~\ref{app:sour}.

Similarly at $n$-th order the $\tau$ equation of motion~\eqref{cons0} becomes 
\be \label{hamn}
\mbox{Tr} \sK_0 \sK_n - K_0 K_n = Y_n
\ee
where the left hand side should be evaluated at zeroth order solution $\tau_0$ and $Y_n$ again depends  only on lower order terms. For example at 2nd order it can be written as 
\be 
 Y_2=  -\frac{1}{2}\dR_2 + \le( \sD^a K_0 - 
\sD_b \sK^{ba}_0 \ri) {\p \tau_0 \ov \p x^a}\ .
 \ee
Note that $\tau_n$ does not appear in~\eqref{hamn} as $\p_u (E_u{^u})_0 =0$.

\subsection{Non-dissipative action at second order?}

We have carried out the evaluation of $I_{\rm UV}$ to second order. The full results are rather complicated 
and will be presented elsewhere.  Here we will only mention results relevant for the following question: 
can we find boundary conditions for $\bar h$ at the horizon which allow us to derive a non-dissipative 
hydrodynamical action to 2nd order in boundary derivatives?  Mathematically this requires that 
in taking $\bar h$ to be null,  $I_{\rm UV} [\bar h, \bar g, X^a]$ should have a well-define limit and 
furthermore $\bar h$ will either decouple (as in the zeroth order) or be determined in terms of $\bar g$ and $X^a$, with $X^a$ {\it unconstrained}. 
There are many reasons not to expect this to happen. After all, the holographic system we are working with has a nonzero shear viscosity, and things will eventually fall into horizon after waiting long enough time. 
Nevertheless it is instructive to work this out explicitly. Note that ideal fluid action of~\cite{Dubovsky:2005xd} has been generalized to second order in derivatives in~\cite{Dubovsky:2011sj,Bhattacharya:2012zx} based on volume-preserving diffeomorphisms. 

For simplicity we will take $\bar g_{\mu \nu} = \eta_{\mu \nu}$. We find that in taking the horizon limit $\De \equiv d(u_h - \tau) \to 0$, $I_{\rm UV}$ develops various levels of divergences in terms of dependence on $\De$: 

\ben 

\item The most divergent terms have the form 
\be 
\sL^{(2)}_{\rm UV} \sim {\tr\log^2\hat \sigma \ov \De^3} \le({1 \ov \log^2 \De} + {1 \ov \log^3 \De} + \cdots \ri)
\ee
where $\hat \sig$ is traceless part of $\sig^\rmi{_\rmj} = \al^{\rmi \rmk} \sig_{\rmk \rmj}$, 
and we have suppressed other \textit{finite} constant coefficients. Interestingly all these divergences go away if we impose 
the regularity condition~\eqref{regcond01} at the horizon which is equivalent to  
\be \label{regcond2} 
\tr\log^2\hat\sigma=0 \ .
\ee

\item The next divergent terms are of the form 
\be 
\sL^{(2)}_{\rm UV} \sim \p_\mu j^\mu \le({1 \ov \De^2} + {1 \ov \De} \ri) 
\ee
where $j^\mu$ is the entropy current~\eqref{encu1}. Thus they vanish if we impose the non-dissipative condition
\be \label{nbs}
\p_\mu j^\mu = 0 \ .
\ee

\item Finally, we have the logarithmic divergence of the form 
\be \label{lindiv1}
\mathcal L^{(2)}=-\frac 1d 2^{3-\frac 4d} e^{(2-d)u_h}\Sigma^2  \log \Delta +O(\Delta^0)
\ee
where
\be \Sigma_{\mu\nu}=P_{\mu\rho}P_{\nu\sigma}\partial^{(\rho}u^{\sigma)}-\frac 1{d-1}\partial_\rho u^\rho P_{\mu\nu},\quad P_{\mu\nu}=\eta_{\mu\nu}+u_\mu u_\nu\ .
\ee
For this divergence to disappear, one then needs
\be \label{shearf}
\Sigma^2=0
\ee
i.e. the system is shear free. Note that the divergence in (\ref{lindiv1}) is proportional to $\Sig^2$, which is precisely the rate of increase of the horizon area.\footnote{To see this explicitly, one need to study the Raychaudhuri equation associated with the null congruence $\ell^a$ on the horizon. In particular, one may need to put on shell the contraction of the Einstein equation with $\ell^a$ at this order. See \cite{Gourgoulhon:2005ng} for details.}

\een 

If we want to have unconstrained $X^a$, the shear-free condition~\eqref{shearf} cannot be consistently imposed. Thus 
it appears not possible to generalize the non-dissipative horizon condition to obtain a second order non-dissipative action. 
As mentioned at the beginning of this subsection, this is hardly surprising. In particular, the specific divergence structure 
of~\eqref{lindiv1} implies that we must take account of dissipation. 

We should note that in the full Schwinger-Keldysh program~\eqref{enne} outlined in the introduction, there is {\it no} need 
to  impose any of the above conditions~\eqref{regcond2},~\eqref{nbs} and~\eqref{shearf}. The divergences will cancel with 
those from $\Psi_{\rm IR}$ after we do a consistent matching at the stretched horizons. Also, the divergences mentioned above are not due to the use of Gaussian normal coordinates, which of course become singular themselves at the horizon. Similar divergences also occur in Eddington-Finkelstein coordinates. Being off-shell means that there are necessarily both in-falling and out-going modes at the horizon (which will further be magnified by nonlinear interactions) which will lead to 
divergences also in the  Eddington-Finkelstein coordinates. 

Finally, for completeness, let us mention that if we do impose all of~\eqref{regcond2},~\eqref{nbs} and~\eqref{shearf}, we obtain a simple result 
\be \label{finact1}
\mathcal L^{(2)}=2^{1-\frac 4d}e^{(2-d)u_h}\left[\frac2{d-1}\theta^2-(d-2)\beta^2-2a^\mu\beta_\mu\right],\ee
where
\be \theta=\partial_\mu u^\mu,\quad \beta_\mu=P_{\mu\nu}\partial^\nu u_h,\quad a^\mu=u^\nu\partial_\nu u^\mu.\ee
The second order Lagrangian (\ref{finact1}) is subject to the ambiguity in the field redefinition
\be \xi^{\tx i}\to \xi^{\tx i}+\delta \xi^{\tx i}
\ee
which we fix by using the zeroth order equation of motion
\be
P^{\mu\nu}\partial_\nu u_h=u^\nu\partial_\nu u^\mu.
\ee
to express $\beta^\mu$ in terms of $a^\mu$. Eq. (\ref{finact1}) can then be simplified to
\be \mathcal L^{(2)}=2^{1-\frac 4d}e^{(2-d)u_h}\left[\frac2{d-1}\theta^2-da^2\right].\ee
We should emphasize that due to various conditions imposed at the horizon, the off-shell nature of the above ``action'' is 
not clear at the moment. To have a genuine off-shell action for one patch we should first compute the full action for both segments of the Schwinger-Keldysh contour, and then integrate out modes of the other patch. At second order this ``integrating-out'' procedure likely does not make sense in the presence of dissipation. Even if this procedure makes sense after suppressing dissipation, it is not clear our current procedure of imposing regularity and non-dissipative conditions yields that.

\section{Conclusion and discussions} \label{sec:diss}

In this paper we outlined a program to obtain an action principle for dissipative hydrodynamics
from holographic Wilsonian RG, and then developed techniques to compute $I_{\rm UV}$, as defined
in~\eqref{uvint1}, at full nonlinear level in the derivative expansion. The ``Goldstone'' degrees of freedom envisioned in~\cite{Nickel:2010pr} arise naturally from gravity path integrals, and the ideal fluid action of~\cite{Dubovsky:2005xd}
emerges at zeroth order in derivative expansion when non-dissipative condition is imposed at the horizon.
The volume-preserving diffeomorphisms of~\cite{Dubovsky:2005xd} appear here as a subgroup of horizon diffeomorphisms.
We also found that a direct generalization of the non-dissipative condition to higher orders does not appear compatible
with the action principle.

An immediate generalization of the results here is to compute $\Psi_{\rm IR}$ of~\eqref{decpl} which will enable us
to take into account of dissipations.

In our discussion we have ignored possible corrections from Jacobian in the change of variables in going from~\eqref{met0} to~\eqref{gnc0}, as well as higher order corrections in the saddle point approximation of gravity path integrals. Such corrections are suppressed at leading order in the large $N$ limit of boundary systems. Nevertheless, they may be important
for understanding the general structure of the hydrodynamical action, thus it would be good to work them out explicitly and explore their physical effects.

It would be interesting to generalize the results to more general situations, such as charged fluids, fluids with more general equations of state (for example~\cite{Kanitscheider:2009as}), systems with
anomalies (such as those considered in~\cite{Erdmenger:2008rm,Banerjee:2008th,Son:2009tf}), 
or higher derivative gravities. 

\vspace{0.2in}   \centerline{\bf{Acknowledgements}} \vspace{0.2in}
We thank A.~Adams, R. Loganayagam, G.~Policastro,  M.~Rangamani,  and D.~T.~Son for conversations, and  M.~Rangamani
for collaboration at the initial stage. Work supported in part by funds provided by the U.S. Department of Energy
(D.O.E.) under cooperative research agreement DE-FG0205ER41360.
We also thank the Galileo Galilei Institute for Theoretical Physics for the hospitality and the INFN for partial support during the completion of this work.

\appendix

\section{Boundary term} \label{app:bd}

\subsection{Boundary compatible with foliation}

Consider a spacetime $M$ with a boundary $\p M$. Suppose $\p M$ is a slice of  a foliation of $M$ by hypersurfaces $\Sig_u$. We denote the outward normal vector to $\Sig_u$ by $n^M$.
The Gauss-Codazzi relation gives
\be \label{r1}
R = {^{(d)}}R +  (K^2 - K_{MN} K^{MN}) - 2 \nab_M (n^M \nab_N n^N -
  n^N \nab_N n^M)
\ee
where ${^{(d)}}R$ is the intrinsic scalar curvature of $\Sig_u$ and $K_{MN}$ is the extrinsic curvature for $\Sig_u$
with
\be
K = \nab_M n^M \ .
\ee
Now let us consider
\be
S = \int_M d^{d+1} x  \sqrt{-g}\,  (R- 2 \Lam) + \int_{\p M} d^d x \sqrt{-h} \, 2 K
\ee
and apply the Stokes theorem to the last term of~\eqref{r1}
\bea
-2 \int_M d^{d+1} x \, \sqrt{-g} \, \nab_M (n^M \nab_N n^N -
  n^N \nab_N n^M) &= & - 2 \int_{\p M} d^d x \, \sqrt{-h} \, n_M (n^M \nab_N n^N -
  n^N \nab_N n^M) \cr
  & = &  - 2 \int_{\p M} d^d x \, \sqrt{-h} \, \nab_M n^M
  \eea
which then directly cancels the Gibbons-Hawking term. In this case we thus find that
\be
S= \int_M d^{d+1} x  \sqrt{-g}\, \le[ {^{(d)}}R +  (K^2 - K_{MN} K^{MN}) \ri] \ .
\ee

\subsection{Boundary incompatible with foliation}

Here we will consider an explicit example with a spacetime metric
\be
ds^2 = du^2 + \ga_{ab} (u, x^a) dx^a dx^b  \ .
\ee
We further consider a foliation of the spacetime by hypersurfaces $\Sig_u$ specified by $u = {\rm const}$.
Denote the normal vector field to $\Sig_u$ by $n^M$, which can be written as
\be
n_M = (1,0), \qquad n^M = (1, 0) \ .
\ee
The extrinsic curvature for $\Sig_u$ can be written as
\be \label{defkk}
K_{ab} = \ha \p_u \ga_{ab}  , \qquad K = \ha \ga^{ab} \p_u \ga_{ab} \ .
\ee
The Gauss-Codazzi relation gives
\be \label{r2}
R = {^{(d)}}R +  (K^2 - K_{MN} K^{MN}) - 2 \nab_M (n^M \nab_N n^N -
  n^N \nab_N n^M)
\ee
where ${^{(d)}}R$ is the intrinsic scalar curvature of $\Sig_u$.
Now suppose the spacetime region $M$ has a boundary $\p M$ which does not coincide with one of $\Sig_u$.  More explicitly, we specify $\p M$ by
\be
 u = \tau (x^a)
 \ee
 for some function $\tau (x^a)$. The outward normal vector to $\p M$ can thus be written as 
 \be
 \ell_M ={ \fn } (1, -\p_a \tau) , \qquad \ell^M =  { \fn} (1, - \p^a \tau) , \qquad \fn = {1 \ov \sqrt{1 + \p_a \tau \p^a \tau}}, \qquad \p^a \tau \equiv \ga^{ab} \p_b \tau \ .
 \ee
The extrinsic curvature of $\p M$ is given by
\bea
K |_{\p M} &= & {1 \ov \sqrt{-\ga}} \p_u (\fn\sqrt{-\ga} ) - {1 \ov \sqrt{-\ga}} \p_a (\fn\sqrt{-\ga}   \p^a \tau) \cr
& = & \fn K + {\fn^3} K_{ab} \p^a \tau \p^b \tau - {\fn} D^2 \tau  + {\fn^3} \p^a \tau \p^b \tau D_a D_b \tau \
\label{onee}
\eea
where we have used~\eqref{defkk} and all indices and covariant derivatives are defined with
respect to $h_{ab} = \ga_{ab} (\tau (x),x^a)$.

The induced metric on $\p M$ is given by
\be
\bar h_{ab} = h_{ab} + \p_a \tau \p_b \tau\ .
\ee
Note the relations
\be
h^{ab} = \bar h^{ab} + {1 \ov \fn^2} \bar \p^a \tau \bar \p^b \tau, \quad \bar \p^a \tau \equiv \bar h^{ab} \p_b \tau
= \fn^2 \p^a \tau, \quad \sqrt{-h} = \fn\sqrt{-\bar h} ,
 \quad \fn =
  \sqrt{1 - \bar \p^a \tau \p_a \tau} \
\ee
and~\eqref{onee} can also be written as
\be
K |_{\p M} = \fn K + {1 \ov \fn} K_{ab} \bar \p^a \tau \bar \p^b \tau - {1 \ov \fn} \bar D^2 \tau  \ .
\ee

Now combining the boundary term in~\eqref{r1} and the Gibbons-Hawking term we find that
\bea
S_{\rm bd} &= & 2 \int_{\p M} d^dx \sqrt{-\bar h} \, K|_{\p M} - 2 \int_{\p M} d^d x \sqrt{-h} \, K \cr
& = & 2 \int_{\sir} d^d x \sqrt{-\bar h} \, {1 \ov \fn} \le( \le( \bar h^{ab} (\bar\p \tau)^2  -  \bar\p^a \tau \bar \p^b \tau  \ri) K_{ab}
-  \bar D^2 \tau \ri) \ .
\eea

\section{Explicitly expressions of sources} \label{app:sour}

Here we give explicit expressions of various sources introduced in Sec.~\ref{sec:high}
\bea
B_n &=& {d-2 \ov 2d} \dR_n - \ha \sum_{i=2}^{n-2} \Tr \KK_i \KK_{n-i} - {d-1 \ov 2d} \sum_{i=2}^{n-2} K_i K_{n-i} \\
P_n & = & \sR_n - {1 \ov d} \dR_n - \sum_{i=2}^{n-2} K_i \KK_{n-i} \\
\label{ssn}
S_n & = & J_n - 3 \sum_{i=2}^{n-2} K_i' K_{n-i} - \sum_{i,j=0'}^{n-2} K_i K_j K_{n-i-j} \\
J_n & = & \sum_{i=0}^{n-2} K_i \dR_{n-i} + {d-2 \ov 2 (d-1)} \dR_n' -{d\ov d-1} \sum_{i=0}^{n-2} \Tr \sK_i \sR_{n-i} \ .
\eea
Note that in the last term of~\eqref{ssn}, the sum should not include the term with $i=j=0$ (which is denoted using a prime).
Also note the relation
\be
S_n = {d \ov d-1} \le(B_n' + 2 K_0 B_n - \Tr \KK_0 P_n \ri) \ .
\ee

\appendix

\bibliography{hydrorefs}

\end{document}